\documentclass[12pt]{article}
\usepackage{cite}
\usepackage{graphicx}
\usepackage{psfrag}
\usepackage{amssymb}
\usepackage{epsfig}

\newcommand \ie {{i.e.}}
\newcommand \eg {{e.g.}}

\makeatletter
\newcommand{\preprint}[1]{%
    \renewcommand{\PreprintNumbers}{%
       \begin{tabular}[t]{l}%
          #1
       \end{tabular}%
    }%
}

\providecommand{\PreprintNumbers}{}

   \def\@maketitle{%
      \newpage
      \null
      \par\noindent\hfill
         \PreprintNumbers
      \par
      \vskip 2em%
      \begin{center}%
         \let \footnote \thanks
         {\LARGE \@title \par}%
         \vskip 1.5em%
         {\large
            \lineskip .5em%
            \begin{tabular}[t]{c}%
               \@author
            \end{tabular}%
            \par
          }%
          \vskip 1em%
          {\large \@date}%
      \end{center}%
      \par
      \vskip 1.5em%
   }
\makeatother
\preprint{%
   DESY--01--014\\
   MSU-HEP-10330\\
   hep-ph/0105207
}
\title{Tests of goodness of fit to multiple data sets}
\author{
   John C. Collins,\footnote{E-mail: collins@phys.psu.edu}
   \footnote{On leave from:
        Physics Department,
        Penn State University, %\\
        104 Davey Laboratory,
        University Park PA 16802,
        U.S.A.
   }
   \\
   DESY, Notkestra{\ss}e 85, D-22603 Hamburg, Germany, \\
   {\em and}\\
   II Institut f{\"u}r Theoretische Physik, Universit{\"a}t Hamburg, \\{}
   Luruper Chaussee 149, D-22761 Hamburg, Germany
\\[8mm]
   Jon Pumplin,\\
        Department of Physics and Astronomy,\\
        Michigan State University,
        East Lansing MI 48824,
        U.S.A.
\\[10mm]
}

\date{21 May 2001}

\begin{document}
\maketitle

\begin{abstract}
We propose a new and rather stringent criterion for testing the goodness
of fit between a theory and experiment.  It
is motivated by the paradox that the criterion on $\chi^2$ for testing a  
theory is much weaker than the criterion for finding the best fit 
value of a parameter in the theory.  We present a method by which the stronger
parameter-fitting criterion can be applied to subsets of data in a
global fit.
\end{abstract}

\newpage

%========================================================
\section{Introduction}

Global fits of theory to large amounts of experimental data are rather
important to current elementary particle phenomenology \cite{CTEQ,MRST,EW}.
A substantial amount of work has been done to estimate the errors on these
fits \cite{Giele:1998gw, Pumplin:2000vx, Stump:2001gu, Pumplin:2001ct},
and it is now of obvious importance to test whether the fits obtained are
actually good.  Is the theory correct? Or is an extension to the standard
model needed?  Are the experiments correct?

The simplest requirement for a good fit
is that the overall $\chi^2$ indicates a good fit
according to the hypothesis-testing criterion, which allows a range
$\sim \! \sqrt{2N}$ in the value of $\chi^2$ (where $N$ is the number of degrees
of freedom).
In fact, as we will show, this criterion is far from optimal.  For
example, a small subset of the data may be quite badly fit, but the
contribution of that subset to the overall $\chi^2$ may nevertheless be
too small for it to show up significantly.

In this paper we propose a new criterion for goodness of fit.  Not only
should the overall $\chi^2$ be good, but the fits to individual experiments
in the data set or to subsets of the data should also be good in a
particular and quite stringent manner.  This leads to a ``parameter-fitting''
criterion that goes beyond the traditional ``hypothesis-testing'' criterion.

\begin{figure}[hbt]
  \centering
  \includegraphics[scale=0.6]{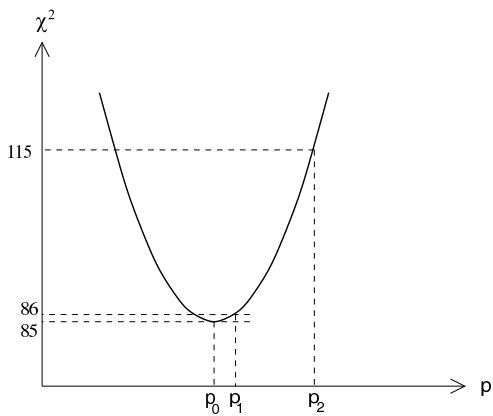}
  \caption{Hypothetical plot of $\chi^2$ vs.\ parameter $p$ in fitting of
    theory to data with $N=100$ points. }
  \label{fig:fitting}
\end{figure}

Our criterion is motivated by the following paradox \cite{paradox}, the
``paradox of parameter determination and hypothesis testing'' which is
illustrated in Fig.\ \ref{fig:fitting}.  If one has a theoretical
prediction for an experiment with $N$ data points, then a good fit should have 
$\chi^2$ approximately in the range $N \pm \sqrt{2N}$, which is the
one-standard-deviation range for $\chi^2$ when the experiment is repeated
many times.
Let us call this the {\it hypothesis-testing} criterion.  On the other hand, if
the theory has a parameter $p$ that is fitted from the data, then the
one-standard-deviation error on that parameter is given by a deviation of
$\chi^2(p)$ by one unit from its minimum.  Let us call this the
{\it parameter-fitting} criterion.  Now observe that if $p$ is varied so as to
give a deviation of $\sqrt{2N}$ of $\chi^2(p)$ from its minimum, it produces 
a large deviation of $(2N)^{1/4}$ standard deviations from the best
fit.\footnote{
   Strictly speaking, $N$ should be replaced by the number of degrees of
   freedom, which in this case is $N-1$.  But for the case of interest,
   this is irrelevant, since $N$ is large. 
}
The paradox is that a particular value of $p$, such as $p_2$, can thus
simultaneously provide a 
good fit according to the hypothesis-testing criterion and a bad fit
according to the parameter-fitting criterion.

\begin{figure}[hbt]
  \centering
  \includegraphics[scale=0.6]{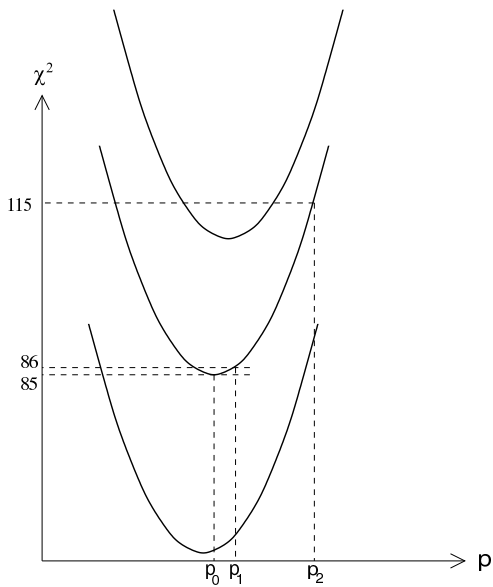}
  \caption{Typical plots of $\chi^2(p)$ when the experiment of 
    Fig.\ \ref{fig:fitting} is repeated.}
  \label{fig:repeat}
\end{figure}

The paradox is resolved by examining what happens if the experiment is
repeated---see Fig.\ \ref{fig:repeat}.  The curves for $\chi^2(p)$
fluctuate 
vertically by a typical amount $\sqrt{2N}$, but horizontally only by a
typical amount corresponding to a one-standard-deviation variation in the
parameter.  If one only knows the predictions of the theory for one
particular value of $p$, then only the weaker hypothesis-testing criterion
($\chi^2$ in the range $N \pm \sqrt{2N}$) can be used.  
But if more information is available, namely the predictions of the theory 
for any value of $p$, then the shape of the $\chi^2(p)$ curve can be
used, and hence a more stringent criterion for goodness of fit is available.

In order to test the goodness of fit to a set of experimental data, one
should obviously examine not only the overall total $\chi^2$, but also the
$\chi^2$ for suitable subsets of data.  Our aim in this paper is to find the
most stringent criterion for doing this.  First, we will introduce the key
idea by means of
a simple example, and we will show that the appropriate criterion is a
version of the parameter-fitting criterion rather than the
weaker hypothesis-testing criterion.
Then we will generalize the resulting criterion to a full
multi-parameter and multi-experiment situation.
Finally we will present a convenient method for
showing the results in one-dimensional plots when there are many
theoretical parameters, and examine those plots for a typical application
that is of current interest.

In addition to the standard concepts of statistical and systematic errors,
we find it useful to introduce the concept of a {\it bug:} an unforeseen
error that is not taken into account in the determination of the systematic
errors, and for which the probability distribution is highly non-Gaussian.
A bug in a
computer program used in the experiment or in the theory calculation is a
canonical example.  But we also find it useful to consider an error in the
theory itself as a bug: an error in the Lagrangian instead of in hardware
or software.  An error in the theory is otherwise known as new physics.  An
actually incorrect experiment is also an example of a bug.

The reason for explicitly introducing the concept of a bug is to suitably
describe what happens when there is an extremely bad fit between theory and
experiment.  In this situation, our usual experience is not that an
extremely improbable fluctuation has occurred in the normal statistical
and systematic effects, but that something has happened that wasn't
allowed for in the estimation of the errors, \ie, a bug has occurred.

The observed distribution of errors found in a study of actual
experiments in particle physics does not in fact follow a Gaussian
form \cite{Bukhvostov}.  Although the peak of the distribution is that of
the expected Gaussian---showing that experimentalists typically estimated
their one-standard-deviation errors correctly---there is a substantial
non-Gaussian tail that one could associate, at least in part, with the
bugs mentioned above.

We see at least two ways to formulate these ideas in terms of a statistical
analysis. The first is simply to formulate an appropriate criterion for
recognizing when a fit is bad.  That is the primary issue addressed in this 
paper.
A second approach, discussed in a separate paper\cite{Pumplin}, is to modify 
the normal formula for $\chi^2$ to take into account the non-Gaussian tails 
of error distributions.  This improves the estimation of parameter values by
allowing the fitting procedure to effectively disregard data points that are 
badly fit 
by the combination of theory and the other experiments.  Such an analysis can 
also be used in a suitably Bayesian sense to deduce the probability of a bug, 
if the non-Gaussian tails are identified with the effects of bugs.

%========================================================
\section{Two experiments; one parameter fit}
\label{sec:TEV.HERA}
We will explain our ideas in their simplest context, by presenting a
hypothetical situation involving the comparison of a theory to two
experiments.

\subsection*{Scenario}

Consider two experiments, which we will call the TEV experiment and the
HERA experiment.  Let the relevant theory be given by standard
perturbative QCD calculations with particular sets of parton densities,
CTEQ and MRST, which have been fit to other data.  Although the numbers we
give are completely hypothetical, we use names representing 
actual experiments and actual parton densities in order to show vividly
that we intend our ideas on statistical methods to be applied to important
practical cases.

\begin{table}[hbt]
  \begin{center}
  \begin{tabular}{||r||c|c|c||}
     \hline
     \hline
              &    TEV      &    HERA      &    Total     \\
      PDF     & (100 points) & (100 points) & (200 points) \\
     \hline
     \hline
      CTEQ    &      85           &       115         &       200         \\
     \hline
      MRST     &     115           &        85         &       200         \\
     \hline
     \hline
  \end{tabular}
  \end{center}
  \caption{Hypothetical values of $\chi^2$ for comparison between theory,
   using two different sets of parton distribution functions (PDF),
   and two sets of experimental data. }
  \label{table:2sets}
\end{table}

Suppose each data set consists of 100 points, and that the values of
$\chi^2$ are as shown in Table \ref{table:2sets}.  Clearly, each set of
parton densities is a good fit to both experiments according to the obvious
criterion, that for hypothesis testing.

But in fact, as we will now show, both sets of parton densities are actually
bad fits to the data.  We will show this by converting the problem to one of
parameter fitting.  Given the pair of parton density sets, any linear
combination of them
\begin{equation}
  \label{eq:f.p}
  f_p = p f^{\rm CTEQ} + (1-p) f^{\rm MRST} 
\end{equation}
is also a valid parton density set.\footnote{ At least if $p$ is not so
  negative or so far above unity that positivity of the parton densities
  is violated.
}
Since the original CTEQ and MRST sets give good fits to
previous data, we should expect that the new parton densities also give
good fits to previous data, if $p$ is in a reasonable range, say $-1$ to
2.

We now ask for the results of a fit for $p$.  Let us hypothesize that the
$\chi^2$ functions for the two experiments are quadratic functions of $p$,
and that they have the following forms, which
reproduce the values in Table \ref{table:2sets}:
\begin{eqnarray}
  \label{eq:chi2-2sets}
  \chi_{\rm TEV}^2 &=& 85 + 30(1-p)^2
\nonumber\\
  \chi_{\rm HERA}^2 &=& 85 + 30p^2 \; .
\end{eqnarray}
The total chi squared is
\begin{equation}
  \label{eq:chi2-total}
    \chi_{\rm tot}^2
    = \chi_{\rm TEV}^2 + \chi_{\rm HERA}^2
    = 185 + 60(p-0.5)^2 \; .
\end{equation}
The best fit has $p = 0.5 \pm 0.13$, and the corresponding best parton density
set is $f^{\rm best} \equiv f_{0.5}\,$.

\subsection*{Global fit in the scenario is not correct}

We now see that the hypothetical CTEQ and MRST parton densities are both about
4 standard deviations from the best fit, and are therefore both strongly
disfavored, as claimed above.  We can obtain the same result by considering
the fits to $p$ that would be performed by the individual experiments.
TEV says that $p = 1.00 \pm 0.18$, while HERA says that $p = 0.00 \pm 0.18$.
These results are inconsistent at the $3.9\,\sigma$ level.

We have moved a long way from the situation apparently given by the
numbers shown in the last column of Table \ref{table:2sets}, where the
hypothesis-testing
criterion says that both the CTEQ and MRST parton densities give good
fits to the data.  By using the extra information that there is a
parameter that can be fitted, we have invoked the much more powerful
parameter-fitting criterion for goodness of fit, and have {\em correctly}
concluded that there is an inconsistency.
The real situation is probably that one (or
both) experiments is wrong, or that the theoretical calculation for one
(or both) experiments is wrong.\footnote{
    When we say that a theoretical calculation might be wrong, we intend
    to encompass a range of possibilities. One is, of course, an
    ordinary calculational mistake; but another is that the theory itself,
    or the approximations used to compute it, 
    might be in error.
}

Since one of the experiments is wrong or has an incorrect theory
calculation, the correct estimate of the value of the parameter is
obtained from the fit to the other experiment.  However, we do not know
which experiment is the culprit.  Thus the correct estimate of the fitted
parameter is {\em not} the value $p = 0.5 \pm 0.13$ obtained from the global
fit.  Rather it is either $p = 0 \pm 0.18$ or $p = 1 \pm 0.18\,$.  Our analysis
cannot tell which of these two values is correct, because it does not tell
us which data or which theoretical calculation
should be discarded.  To proceed in this case, 
we must investigate each experiment and its theory to discover what is wrong.
In the meantime, we would have to make do with the range
$p = 0.50 \pm 0.68$ that includes both experiments.

Even that extended range may not include the true situation, since an
important possibility is that both experiments are incompatible with
the theory, in which case the whole global fit to the TEV and HERA data
is inapplicable.  This situation could arise, for example, if there is
some new physics that is important for both of the new experiments but which is
not accessible to the earlier experiments on which the CTEQ and MRST parton
densities were based.

The particular values of $p$ that are preferred depend on the precise form
for the $\chi^2$ for each experiment, which was hypothesized in Eq.\
(\ref{eq:chi2-2sets}).  However, the fact that the experiments are
inconsistent and that they prefer significantly different values of $p$
depends only on the values of $\chi^2$ in Table \ref{table:2sets}.  That is,
we do not actually need to do the parameter fitting to see that there is a
problem.  It is enough to show that the $\chi^2$ for one experiment can be
reduced by many units in going from one parton density set to another,
while the total $\chi^2$ for all of the experiments increases by only
a small amount.  According to the parameter-fitting criterion, there are
therefore two parton density sets, each of which is strongly preferred over
the other, by different sets of data.

\subsection*{Consequences of inconsistency}

After deducing that an experiment is inconsistent with a theory calculation,
we must try to figure out what went wrong.  There is the usual list of
suspects, including:
\begin{itemize}
\item An error in the experiment (\eg, a bug in the data analysis
  software), or any other kind of error in the experiment (\eg, an
  experiment that is simply wrong due to an unforeseen background or
  a mis-measured target size, to recall actual instances).
\item A technical error in the theoretical calculations (\eg, a bug in
  software, or a QCD calculation taken to insufficiently high order in
  perturbation theory). 
\item New physics (\ie, an error in the Lagrangian used as a basis for
  the theory calculations).
\end{itemize}
As suggested before, we will label all of these as bugs, which are
defined as infrequently occurring errors that were not allowed for
when the systematic errors for the experiment and theory were estimated.

The kind of error that we call a bug can produce large effects on the
cross section or on the theory calculation, so the distribution of
effects due to bugs, considered over many experiments, is strongly
non-Gaussian.  Thus the estimate of probabilities from the quadratic
approximation to $\chi^2$ is badly wrong for them.  Since there is a
single large source of error in such cases, we cannot appeal to
the Central Limit Theorem to expect a Gaussian distribution.

Our analysis of $\chi^2$ for the experiments as a function of the parameter
$p$ cannot tell us where the bug is.  It merely tells us when it is
likely that there is one.  Given our previous experience with science, we
know that the probability of a bug is non-negligible.  Indeed, if we
identify bugs with the non-Gaussian component of the distribution of
experimental errors, then Bukhvostov's results \cite{Bukhvostov} imply
that the probability of bugs in a certain class of high-energy physics
experiments is about 15\%!   Not all of these bugs are readily
identifiable.  The identifiable bugs are those in the
strongly non-Gaussian tail of the error-distribution.  For example,
Bukhvostov observes
that 65 data points out of 933 have a deviation greater than $3\,\sigma$.
That corresponds to about $7\%$ of the data, whereas a Gaussian
distribution would predict only $0.3\%$.

Given a particular scenario, we have deduced that is likely (in an
appropriately Bayesian sense) that there is a bug.  Hence it is a correct
scientific decision to investigate the experiments and the theory to
locate the bugs.  The issue for us now is how to quantify this decision more
generally.

%========================================================
\section{General case}

In the previous section, we were able to diagnose that certain parton
densities gave a bad fit to data because of the existence of two different
sets of parton densities.  Now we must ask how CTEQ could find the problem
without MRST's assistance (or vice versa).  One method is to pick a
significant parameter and to examine the dependence of $\chi^2$ on that 
parameter for particular experiments.  An example of this for the MRST parton
densities is given by Fig.\ 21 in Ref.\ \cite{MRST}, where $\chi^2$ for
different experiments is plotted against $\alpha_s(M_Z^2)$.  In the scenario of
the previous section, the comparison of two sets of parton densities
also focused our attention on a different particular parameter.

But in a typical global fit, there are many parameters.  So the issue we
now address is how to automatically find the optimal combinations of
parameters for detecting a bad fit.

We therefore consider a general situation in which we have many data
points and 
experiments, and many parameters in the theory.  We will choose ahead of
time to divide the data into subsets.  These subsets could be individual
experiments, or data points obtained using similar experimental techniques,
or data that rely on a specific aspect of theory.  An example would be
jet data from a particular experiment, possibly divided into regions of low,
medium and high transverse energy.  The idea is to choose subsets of data
that are likely to be simultaneously affected by a typical bug.  Let there
be $g$ subsets (or groups) of data.  The total $\chi^2$, which is a function 
of the theory parameters ${\bf p}$, is the sum of the $\chi^2$ for the 
individual subsets:
\begin{equation}
  \label{eq:total}
  \chi_{\rm tot}^2({\bf p}) = \sum_{i=1}^g \chi_i^2({\bf p}) \; .
\end{equation}
If necessary, the formula for $\chi^2$ may be fudged to take into account
badly estimated correlated systematic errors, as is common practice in 
global fits for parton densities \cite{Stump:2001gu}.

We want to ask whether the fit is improbable at some level, $c\%$. We might
choose $c\% = 5\%$ or even $c\% = 10\%$ as the level below which further
investigation is warranted.  Our proposal is as follows:
\begin{enumerate}
\item Apply normal methods to find the best fit, with
   ${\bf p} = {\bf p}_{\rm best}$ defined by the minimum of
   $\chi_{\rm tot}^2({\bf p})$.
\item If $\chi_{\rm tot}^2({\bf p}_{\rm best})$ is too high according to the
   hypothesis-testing criterion, with $P<c\%$, then we have a bad fit.
\item Similarly if for one or more of the individual experiments,
  $\chi_i^2({\bf p}_{\rm  best})$ is too high according to the
  hypothesis-testing 
  criterion, then again
  we have a bad fit.  (In the case that only one experiment is a bad fit,
  the border line to declare a bad fit is more stringent than for the overall
  fit:  The probability would have to satisfy $P<c\%/g$ to be considered a bad
  fit, because the bad fit could have occurred in any of $g$ places.)
\item Now let us define the region of an overall good fit as the region
  where $\chi_{\rm tot}^2({\bf p})-\chi_{\rm tot}^2({\bf p}_{\rm best})$ is
  less than
  about $\sqrt{2N}$, where $N$ is the total number of data points.  It is
  not necessary to be too precise about this region. It only forms a basis
  for further exploration of goodness or badness of fit.  One does not
  want to investigate values of the parameters that are much outside this
  region, because they then give an unambiguously bad fit.  In
  addition, one can exclude parameter values that are known on other grounds
  to be physically wrong or implausible.
\item Find the minimum of each $\chi_i^2({\bf p})$, for the subsets of data,
  when the parameters range over the region just defined.
  (This is easily done by using the method of Lagrange multipliers.) Let the
  resulting minimum values be $\chi_{i,{\rm min}}^2$.
\item Now compute the difference between the $\chi_i^2$ at the best global
  fit and the minimum that was just calculated, \ie, $\chi_i^2({\bf p}_{\rm
    best}) - \chi_{i,{\rm min}}^2\,$.  If one or more of these is above a
  threshold for a bad fit, {\em in the sense of parameter fitting}, then
  the fit is bad.
\end{enumerate}
Steps 5 and 6 are the novel parts of our proposal.  A possible variation
on these steps, based on mapping the variation of $\chi_i^2$ with
$\chi_{\rm tot}^2$, is described in Sec.\ \ref{sec:CTEQ5app} and
Appendix \ref{sec:one.param.model}. 
Whenever it is determined in one of these ways that a fit is bad, then further
investigation is called for to attempt to discover the reason.
Several caveats are in order:
\begin{itemize}
\item If a particular subset contains very few points and there are many
  parameters, it may be possible to get a $\chi_i^2$ much less than the number
  of points simply because there are many parameters.  Typically, however,
  any particular subset of data determines only a few parameters; perhaps
  only one.  We do not address here how to determine the relevant number
  of parameters or to determine what effect that has on our criterion.
\item A literal use of our criterion requires that the error estimates be
  valid. In particular the correct correlated systematic errors must be
  used.  However, it is common that properly correlated systematic errors
  are not available for experiments, and in that case some appropriate
  allowance must be made.
\end{itemize}

%========================================================
\section{Presentation of results in one-dimensional graphs}

The exploration of the parameter space in many dimensions is difficult to 
visualize.  One way to study 
it is to select a particularly significant parameter and plot
the $\chi_i^2$ for each experiment as a function of that parameter,
while the other parameters are continuously adjusted to optimize the fit.
Examples of such plots are to be found in Refs.\
\cite{MRST,Stump:2001gu}. 

However, this procedure is really only useful when one has identified
a particularly significant direction in parameter space.  So we now
propose a more general way to plot the results.  In fact, we propose two
ways to make the plot, because it is unclear to us at this stage which form
will be more useful.

%------------------------------------------------
\subsection{Plot of $\chi_i^2$ against $\chi_{\rm tot}^2$}
\label{sec:Plot}

In the first method, for each value of $\chi_{\rm tot}^2$ we plot the
minimum of $\chi_i^2({\bf p})$ that is compatible with that value of
$\chi_{\rm tot}^2$.  We thus obtain curves of the $\chi_i^2$ for each
particular experiment against the total $\chi^2$.  One can read off
from the plots how well these experiments agree with the overall global
fit.

\begin{figure}[hbt]
  \centering
  \psfrag{chi2i}{$\chi_i^2$}
  \psfrag{chi2min}{$\chi_{\rm min}^2$}
  \psfrag{chi2total}{$\chi_{\rm tot}^2$}
  \psfrag{A}{A}
  \psfrag{B}{B}
  \psfrag{C}{C}
  \psfrag{D}{D}
  \includegraphics[scale=0.7]{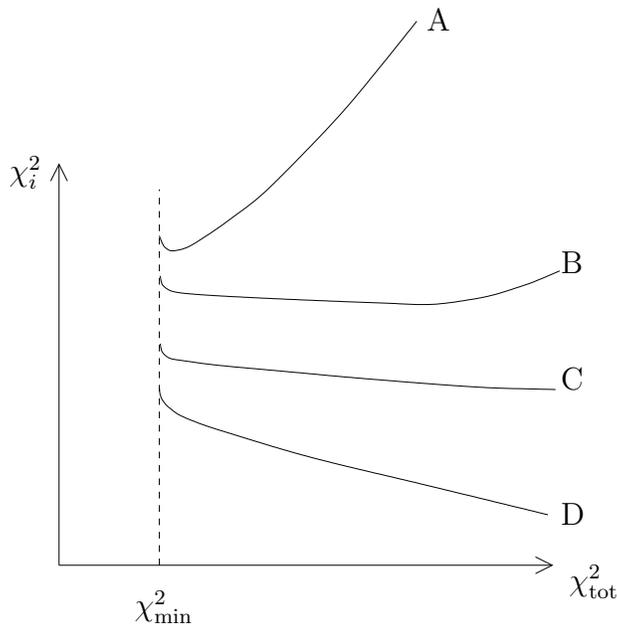}
  \caption{Possible results of plotting $\chi_i^2$ for subsets of data as
    a function of $\chi_{\rm tot}^2$.}
  \label{fig:chi2-chi2}
\end{figure}

Rough sketches of hypothetical examples of such plots are shown in
Fig.\ \ref{fig:chi2-chi2}.  Curve A corresponds to an experiment that
agrees with the global fit and strongly determines all of the parameters.
Curves B and C correspond to experiments that agree with the global fit,
but that do not determine all the parameters.  Finally, curve D corresponds
to an experiment that is in disagreement with the global fit.  (Of course,
this analysis cannot determine whether it is experiment D, one of the other
experiments, or the theory that is in error.)
The criterion that an experiment disagrees with the global fit is that
its $\chi^2$ decreases by more than the amount allowed by the
parameter-fitting criterion.

These curves are straightforward to compute by the Lagrange multiplier
method (cf.\ \cite{Stump:2001gu}).  One minimizes
\begin{equation}
  \label{eq:lag.mult}
  f_\lambda({\bf p}) \, = \, (\lambda - 1) \, \chi_i^2({\bf p})
  \, +  \, \chi_{\rm tot}^2({\bf p})
\end{equation}
for various values of the parameter $\lambda$.  Then for each value of
$\lambda$, the corresponding values of $\chi_i^2$ and $\chi_{\rm tot}^2$ give
one point on the graph.  Performing the minimization of (\ref{eq:lag.mult})
gives a parametric representation of the curve of minimum $\chi_i^2$ against
$\chi_{\rm tot}^2$.

Curves such as those in Fig.\ \ref{fig:chi2-chi2} would be generated using
$\lambda > 1$ in (\ref{eq:lag.mult}), so that experiment $i$ is weighted more
heavily than in the Best Fit, and hence its $\chi_i^2$ is reduced relative to
its value at the
global minimum.  Meanwhile, it is also useful to minimize $f_\lambda({\bf p})$
using $\lambda$ in the range $0 < \lambda < 1$,
since that reveals how the fit to all
the other experiments (as measured by $\chi_{\rm tot}^2 - \chi_i^2$) can be
improved when the fit to experiment $i$ is allowed to get worse (as measured
by an increase in $\chi_i^2$).  This region can be included in graphs like
those illustrated in Fig.\ \ref{fig:chi2-chi2}, where it adds an additional
branch to each curve, beginning at $\chi_{\rm min}^2$; but we defer it instead
until Sec.\ \ref{sec:SecondMethod}.

\subsubsection*{General expectation: monotonic decrease is normal}

As we will now explain, a curve like B is unlikely.  In that curve, the 
$\chi_i^2$ for
the experiment decreases slightly and then increases again as
$\chi_{\rm tot}^2$ increases.  The most common situation is probably curve
C,
where the $\chi_i^2$ for the experiment decreases slightly with the global
$\chi_{\rm tot}^2$; and never rises again, at least not in the relevant
range of $\chi_{\rm tot}^2$.

The reason for this is that one experiment normally only determines a fraction
of the parameters in a global fit.  For example, a neutrino DIS experiment
with a limited range in $Q^2$ tells us a lot about the flavor-separated
quark and antiquark densities, but says relatively little about the
gluon density and the value of $\alpha_s$.  On the other hand, a jet production
experiment at a hadron collider constrains the gluon density and $\alpha_s$,
but does little to discriminate the flavors of quarks and antiquarks.

If we choose a value of $\chi_{\rm tot}^2$ that is just a small amount
above the minimum, then
only a limited range of parameters is allowed.  As we increase the
chosen value of $\chi_{\rm tot}^2$, the range for the parameters increases.
This implies, for example, that the parameters for the quark densities can
be adjusted to give a better fit to the neutrino experiment.  At the same
time, the large value of $\chi_{\rm tot}^2$ is maintained by having a poor
form for the gluon density.  Since the bad gluon density gives hardly any
contribution to the $\chi_i^2$ for the neutrino experiment, the $\chi_i^2$ for
the neutrino experiment will decrease as $\chi_{\rm tot}^2$ increases.

This situation is general,
since it is normal that individual experiments and subsets of data 
strongly determine only a subset of parameters.  Thus the curves in Fig.\
\ref{fig:chi2-chi2} will commonly fall monotonically with $\chi_{\rm tot}^2$,
as in curves C and D.  A rising curve like A or B will
only occur for a subset of data that significantly constrains all of the
parameters.  At the boundary where $\chi_{\rm tot}^2$ approaches its minimum
value, the curves always become vertical, since the derivative of
$\chi_{\rm tot}^2$ with respect to any parameter must be zero at the minimum.

Note that a mildly varying curve like B or C is quite normal in a good
fit.  It indicates that the experiment or subset of data in question is
compatible with the global fit but that it does not determine all the
parameters.  The subset of data might nevertheless be the most significant
in determining some subset of the parameters.

%------------------------------------------------
\subsection{Plot of $\chi_i^2$ against $\chi_{{\rm not\,}i}^2$}
\label{sec:SecondMethod}

A second method to visualize the character of the global fit 
is to plot the minimum of $\chi_i^2$ not against the total
$\chi^2$, but against the $\chi^2$ for the remaining data, \ie{} against
\begin{equation}
  \label{eq:chi2.other}
  \chi_{{\rm not\,}i}^2 \equiv \chi_{\rm tot}^2 - \chi_i^2 =
  \sum_{j \neq i} \chi_j^2 \; .
\end{equation}
This has the advantage of relating two independent contributions to the
total $\chi^2$, and gives the plot some simple but interesting mathematical
properties in a neighborhood of the overall best fit.

\begin{figure}[hbt]
  \centering
  \includegraphics[scale=0.6]{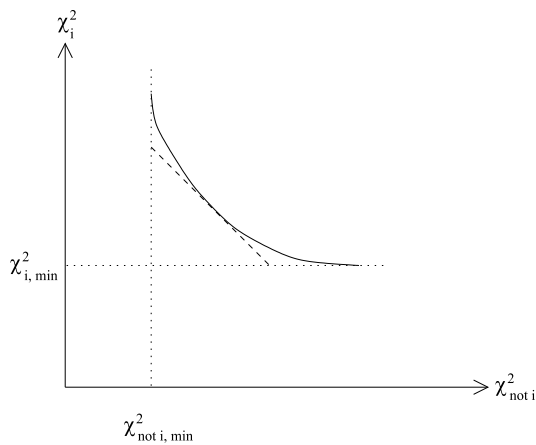}
  \caption{Possible results of plotting the minimum of $\chi_i^2$ for a
      subset of the data as a function of $\chi^2$ for the remaining data.
      The diagonal dashed line at $-45^\circ$ is tangent to the curve at
      the point where $\chi_{\rm tot}^2$ has its minimum value.}
  \label{fig:chi2-i-other}
\end{figure}

This curve can be extracted from the same Lagrange multiplier results
that were used in the previous method, since the quantity
(\ref{eq:lag.mult}) that was minimized there can be written as
\begin{equation}
  \label{eq:lag.mult2}
   f_\lambda({\bf p}) = \lambda \chi_i^2({\bf p}) 
                        + \chi_{{\rm not\,}i}^2({\bf p}) \; .
\end{equation}
Minimizing $f_\lambda$ with respect to the theory parameters ${\bf p}$ gives the
minimum of $\chi_i^2$ for some value of $\chi_{{\rm not\,}i}^2({\bf p})$.
Varying $\lambda$ allows one to plot the minimum $\chi^2$ for experiment
$i$ against $\chi^2$ for the other experiments; \ie, the curve is again
defined parametrically as a function of $\lambda$. A typical form to be
expected for the plot is shown in Fig.\ \ref{fig:chi2-i-other}.

Let ${\bf p}(\lambda)$ be the position of the minimum of the function
(\ref{eq:lag.mult2}).  The requirement of a minimum implies that for all
small variations of the parameters ${\bf p}$ about ${\bf p}(\lambda)$, the
variations of the two components of $\chi^2$ satisfy
\begin{equation}
  \lambda \, \delta\chi_i^2 + \delta\chi_{{\rm not\,}i}^2 = 0 \; ,
\end{equation}
and hence on the curve being plotted
\begin{equation}
\label{eq:gradient}
  \frac{ d\chi_i^2 }{ d\chi_{{\rm not\,}i}^2 } = - \frac{ 1 }{ \lambda } \; .
\end{equation}
It follows that the curve always has the qualitative shape shown in Fig.\ 
\ref{fig:chi2-i-other}.  The best overall fit corresponds to the point $\lambda
= 1$, for which the quantity being minimized is just $\chi_{\rm tot}^2$.
This point corresponds to the geometrical situation shown on the graph,
where a $-45^\circ$ line is tangent to the curve relating the two $\chi^2$s, as
indicated by the dashed line.  The portion of the curve to the right of
the best fit is generated by $\lambda > 1$, and carries the same information as
Fig.\ \ref{fig:chi2-chi2}.  The region to the left is generated by $0 < \lambda
< 1$, where experiment $i$ is de-emphasized in the fit, so that the other
experiments are fit better while experiment $i$ is fit worse.

Figure\ \ref{fig:chi2-i-other} expresses a situation that is essentially the
same as in our original scenario of the TEV and HERA data in
Sec.\ \ref{sec:TEV.HERA}.  The roles
of the TEV and HERA experiments are played by experiment $i$ and
all-the-other-experiments.  Exactly the same criterion for consistency
should be applied:  We have an inconsistency if, at the best global fit,
$\chi_{\rm tot}^2$ exceeds the sum of the
absolute minima for the two subsets of data by more than a few units (the
parameter-fitting tolerance), \ie, if
\begin{equation}
  \label{eq:criterion}
  {\rm min}(\chi_{\rm tot}^2)
  - {\rm min}(\chi_i^2)
  - {\rm min}(\chi_{{\rm not\,}i}^2)
  > \mbox{parameter-fitting tolerance} .
\end{equation}
Of course, the same plot should also be made with experiment $i$ replaced
in turn by each of the other experiments.

%========================================================
\section{Application: CTEQ5}
\label{sec:CTEQ5app}

\begin{figure}[hbt]
  \centering
  \includegraphics[scale=0.75]{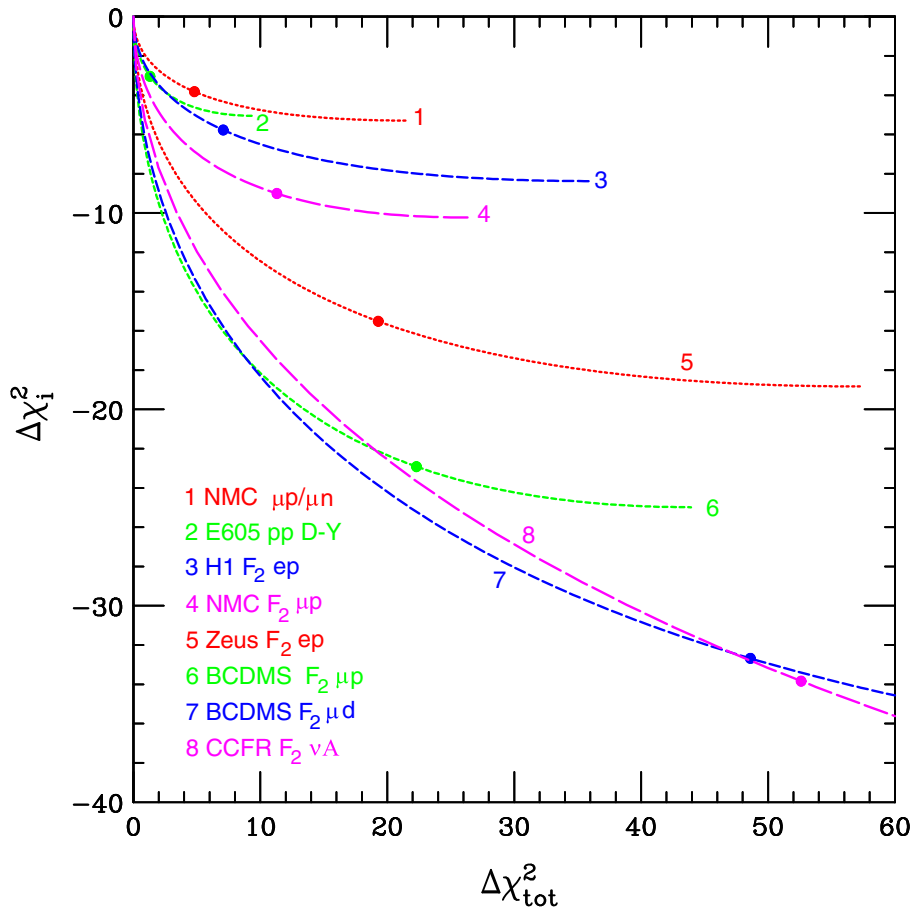}
  \caption{Variation of $\chi_i^2$ with $\chi_{\rm tot}^2$ for 8 of the
   data sets 
   of the CTEQ5 parton density analysis.  Dots mark the points found by
   $\lambda = 5$ in Eq.~(\ref{eq:lag.mult}).
  }
  \label{fig:fit30a}
\end{figure}

As a first practical application of the ideas presented here, we have examined 
the CTEQ5 parton distribution analysis \cite{CTEQ}.  Fig.\ \ref{fig:fit30a}
shows a realization of the generic Fig.\ \ref{fig:chi2-chi2} for the 8
experimental data sets that contribute the lion's share of data points 
($1115$ out of $1295$) to that analysis.  The data sets are numbered in 
Fig.\ \ref{fig:chi2-chi2} in the order of decreasing consistency with the rest 
of the global analysis, as will be shown in Table \ref{table:cteq}.
We subtract the best-fit values, and 
therefore plot
\begin{equation}
  \Delta \chi_i^2 = \chi_i^2 - \chi_i^2(0)
  \label{eq:delchiisq}
\end{equation}
versus
\begin{equation}
  \Delta \chi_{\rm tot}^2 = \chi_{\rm tot}^2 - \chi_{\rm tot}^2(0) \; ,
  \label{eq:delchitsq}
\end{equation}
where the argument $(0)$ denotes values at the minimum of $\chi_{\rm tot}^2$.

Fig.\ \ref{fig:fit30a} shows that for several of the data sets, $\chi_i^2$
can decrease by many units within the range of parameters for which
$\chi_{\rm tot}^2$ increases by $\sqrt{2N} \simeq 50$.  We conclude that the
combined CTEQ5 data set is therefore {\it not} internally consistent
according to the parameter-fitting criterion---even if we make a
substantial allowance for the neglect of correlations among the errors
used to define $\chi_i^2$. (It is also necessary to make an estimate of the
expected decrease in $\chi_i^2$ given the total number of parameters and the
degree to which they are determined by the experiment in question. But we do
not do this here.)

\begin{figure}[hbt]
  \centering
\centerline{
\epsfxsize=0.50\textwidth \epsfbox{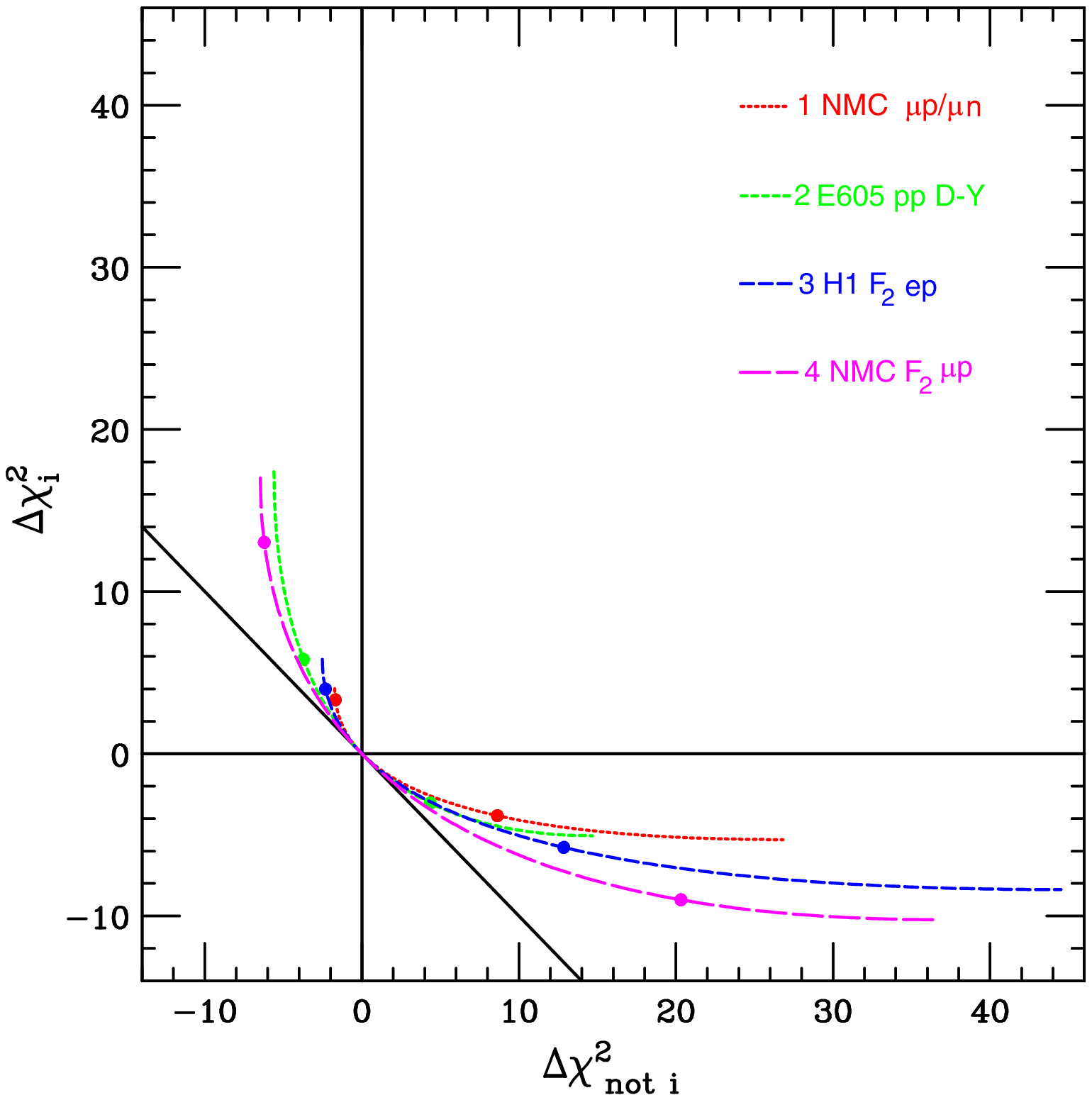}
\hfil
\epsfxsize=0.50\textwidth \epsfbox{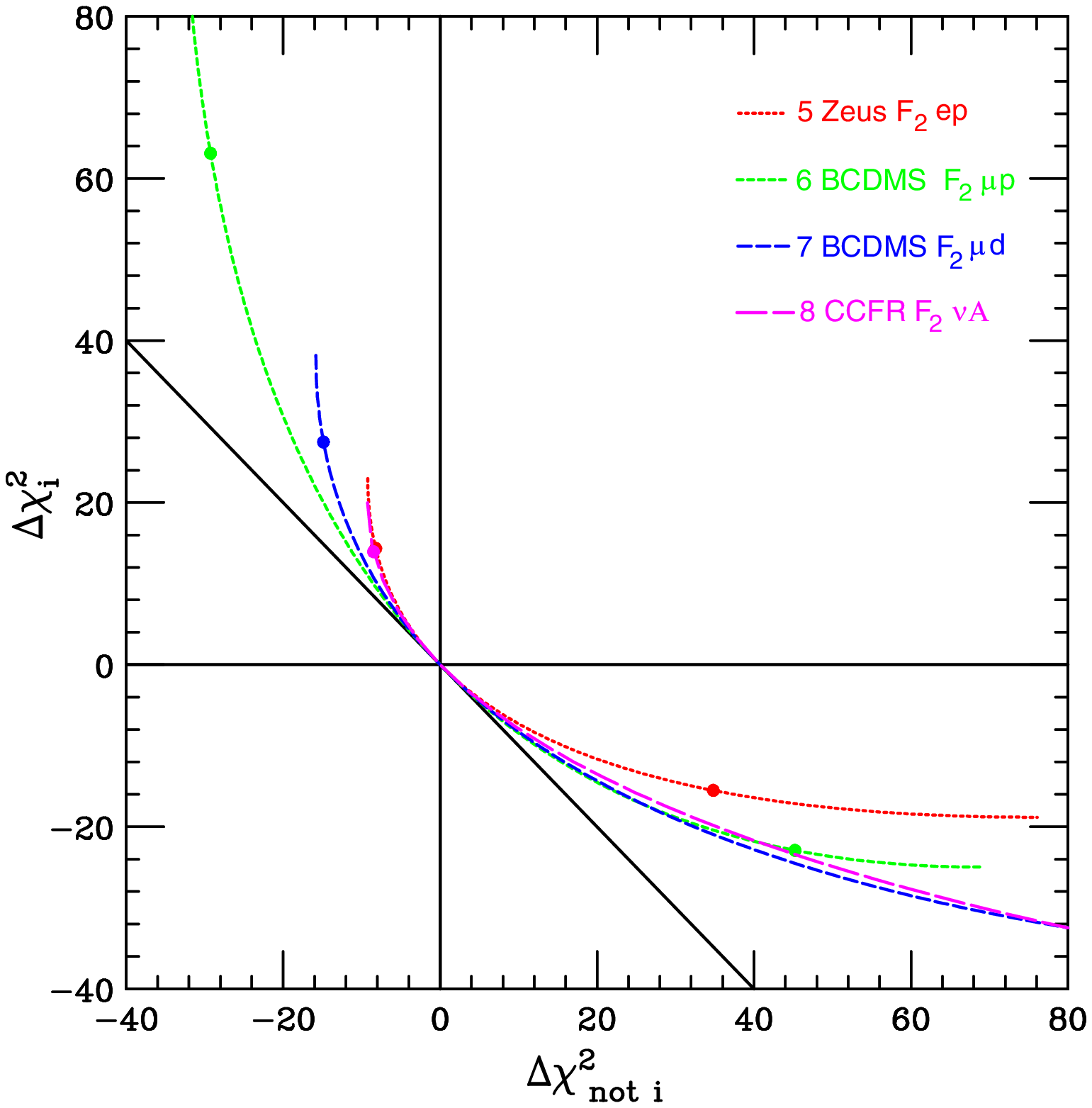}
}
  \caption{Variation of $\chi_i^2$ with $\chi_{{\rm not\,}i}^2$ for 8
   data subsets of the CTEQ5 parton density analysis.  Dots mark the points
   found by $\lambda = 5$ for $\Delta\chi_i^2 < 0$, and $\lambda = 0.2$ for
   $\Delta\chi_i^2 > 0\,$.}
  \label{fig:fit30b}
\end{figure}

Fig.\ \ref{fig:fit30b} similarly shows a realization of the generic
Fig.\ \ref{fig:chi2-i-other} for the same 8 data sets.  The inconsistency
between theory and these data sets according to Steps 5 and 6 of our
parameter-fitting criterion is again apparent.

The curves in Figs.\ \ref{fig:fit30a} and \ref{fig:fit30b} were obtained by
minimizing $f_\lambda({\bf p})$ of (\ref{eq:lag.mult}) with respect to the 16 
free parameters
${\bf p}$ of the CTEQ5 fit, using approximately 18 values of the Lagrange
multiplier parameter $\lambda$ for each data set $i$.  As an aid to plotting
smooth curves, these results were fitted to a simple two-parameter model that
is described in Appendix \ref{sec:one.param.model}.  That model was found to
provide a good description 
of the variations of $\chi_i^2$ with $\chi_{{\rm not\,}i}^2$, while serving to
smooth over small variations and numerical effects that are unimportant for our
purposes.

The model of Appendix \ref{sec:one.param.model} also provides a direct measure 
of the internal consistency of the data sets.  Specifically, its parameter $S$ 
measures the number of standard
deviations by which the value of an effective parameter $q$, as measured
by data set $i$, differs from its value as measured by the combination of all
the other data sets.
(Parameter $q$ represents the combination of fit parameters that data set $i$
is most sensitive to, in conflict with the other data sets.  
It is therefore a different---and generally
nonlinear---function of the actual fit parameters for each data set.)

The fitted values of parameter $S$ are listed in Table \ref{table:cteq}.
We see that many of these data sets are distinctly inconsistent with the 
rest of the data, since the parameter $S$ is considerably larger than
$1.0\,$.  Meanwhile, the parameter $\tan\phi$ is generally less than
$1.0$, which
indicates that each data set (with the exception of set $6$), is
somewhat 
less effective in determining its parameter $q$ than is the remainder of the
data sets put together.

The inconsistency of the data sets used for global fits to parton
density has been known to practitioners for some time.  It was 
quantified in \cite{Pumplin:2001ct} by a simple means of finding the decrease
in $\chi_{{\rm not\,}i}^2$ that could be produced by removing data set $i$
from the fit.  In our notation, that corresponds to the point $\lambda = 0$,
which provides the asymptote of minimum $\Delta\chi_{{\rm not\,}i}^2$ in
Fig.\ \ref{fig:fit30b}.\footnote{The results given in \cite{Pumplin:2001ct} 
for the $\lambda=0$ shifts are somewhat different from those shown in 
Fig.\ \ref{fig:fit30b} because, for conceptual simplicity, we have defined 
$\chi^2$ using weights $1.0$ for all experiments, rather than using the 
CTEQ5 choices.}

% results from /fitcollins/fit30.f (3/20/01)
% expt6.dat    chisq(0)= 1225.781 chisq1(0)= 112.731 S=   2.5990 tanphi= 0.56
% expt10.dat   chisq(0)= 1225.951 chisq1(0)=  93.837 S=   2.8751 tanphi= 0.45
% expt3.dat    chisq(0)= 1225.730 chisq1(0)= 108.282 S=   3.2974 tanphi= 0.58
% expt5.dat    chisq(0)= 1226.536 chisq1(0)= 108.312 S=   4.3356 tanphi= 0.78
% expt4.dat    chisq(0)= 1225.848 chisq1(0)= 247.355 S=   5.3451 tanphi= 0.63
% expt1.dat    chisq(0)= 1226.310 chisq1(0)= 153.307 S=   7.5656 tanphi= 1.04
% expt2.dat    chisq(0)= 1225.896 chisq1(0)= 181.154 S=   7.7708 tanphi= 0.56
% expt8.dat    chisq(0)= 1225.870 chisq1(0)=  73.403 S=   8.1260 tanphi= 0.39
% expt501.dat  chisq(0)= 1079.860 chisq1(0)= 134.286 S=   7.0438 tanphi= 1.17
% expt502.dat  chisq(0)= 1079.426 chisq1(0)= 154.551 S=   7.2139 tanphi= 0.54
% expt503.dat  chisq(0)= 1079.526 chisq1(0)=  96.789 S=   3.0550 tanphi= 0.55

\begin{table}
  \begin{center}
  \begin{tabular}{||c||c|c|c|c|c|c|c|c||}
 \hline
 \hline
Expt       & $1$   & $2$   & $3$   & $4$   & $5$   & $6$   & $7$   & $8$ \\
 \hline
 \hline
$S$        &$2.7$  &$3.3$  &$3.3$  &$4.2$  &$5.3$  &$7.6$  &$7.4$  &$8.3$ \\
 \hline
$\tan\phi$ &$0.56$ &$0.54$ &$0.99$ &$0.86$ &$0.71$ &$1.14$ &$0.65$ &$0.39$ \\
 \hline
 \hline
  \end{tabular}
  \end{center}
  \caption{Fits of Eqs.\ (\ref{eq:delchi1sq})--(\ref{eq:delchi2sq})
    for 8 of the experiments in the CTEQ5 analysis (numbered as
    in Figs.\ \ref{fig:fit30a} and \ref{fig:fit30b}).}
  \label{table:cteq}
\end{table}

What to do about this situation---other than the long-term option of
waiting for the discrepancies to be resolved by improvements in theory or
experiment---remains an open question.  In order to obtain provisional
results in the interim, Ref.\ \cite{Pumplin:2001ct} advocated estimating
the uncertainty of predictions based on the global fit as those contained
in the region $\Delta\chi_{\rm tot}^2 < T^2$, where $T \approx 10$ was 
chosen in order to assume a range of uncertainty for the predictions of 
the global fit that is somewhat broader than the variations of the 
experiments going into it, which are indicated in 
Table\ \ref{table:cteq}.  Of course, any estimate of the uncertainties based 
on an inconsistent fit will necessarily be somewhat model-dependent.

There is reason to hope that at least one discrepancy will be resolved in
the near future by new experimental data.  That is between data sets $3$
and $5$, which correspond to data on $F_2$ from the two major HERA
experiments H1 and ZEUS.  These are similar physical measurements made by
similar techniques, and yet we find that improving the fit to either of
these two experiments (by emphasizing it in the fit with an appropriate 
Lagrange factor $\lambda$) makes the fit to the other experiment get worse.  
This suggests that the problem is an
experimental error that may be resolved in the new data that is expected
soon from these groups.

The largest values of the inconsistency parameter $S$ in Table
\ref{table:cteq} are produced by data sets $7$ and $8$, which are data from
muon scattering on deuterium and neutrino scattering on a nuclear target.
One may speculate that the problems are caused by inaccurate treatment of 
nuclear binding effects in these experiments, although the discrepancy 
associated with the $\mu d$ data is not much worse that associated with the 
$\mu p$ data measured by the same group.

{\em One should be cautious in interpreting the value of $S$ as a definite
number of standard deviations, since the interpretation derived in Appendix
\ref{sec:one.param.model} assumes that only one effective parameter is 
determined by
a particular experiment.} 

For the inconsistency to be a real effect, it should be confirmable in
another global fit.  That this is the case for the MRST fit can be seen in
Fig.\ 21 of \cite{MRST}.  There $\chi^2$ is plotted as a function of
$\alpha_s(M_Z^2)$ for several experiments.  The CCFR $F_2$ and the BCDMS
$F_2^{\mu p}$ are clearly very inconsistent with the MRST fit, in agreement
with our results.  We also find a strong inconsistency with the BCDMS
$F_2^{\mu d}$ data, but that data does not appear in the MRST plot.

\begin{figure}
  \centering
  \includegraphics[scale=0.90]{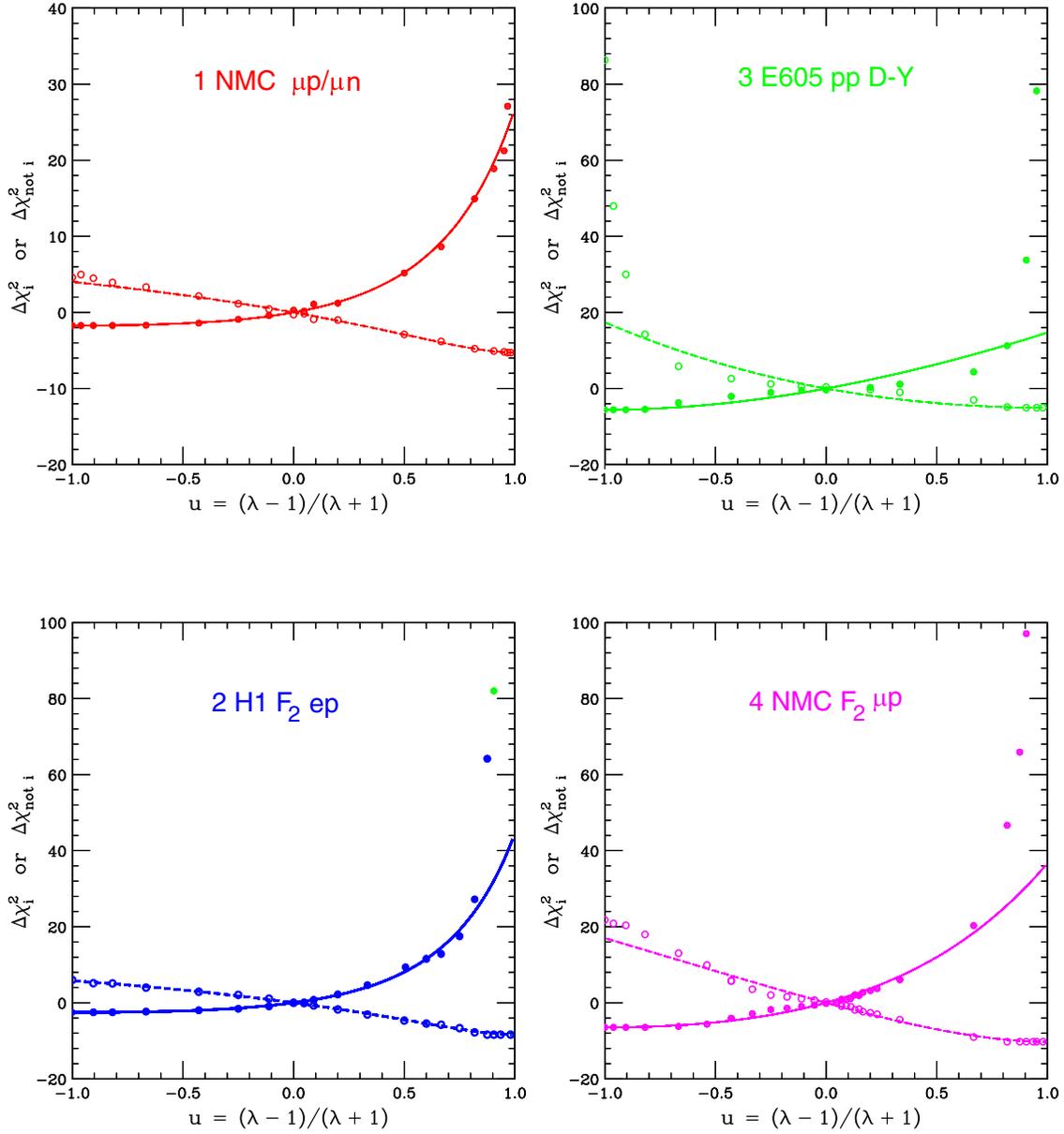}
  \caption{Variation of $\chi_i^2$ (solid curves) and $\chi_{{\rm not\,}i}^2$ 
   for experiments 1--4.
  }
  \label{fig:fit301234}
\end{figure}

\begin{figure}
  \centering
  \includegraphics[scale=0.90]{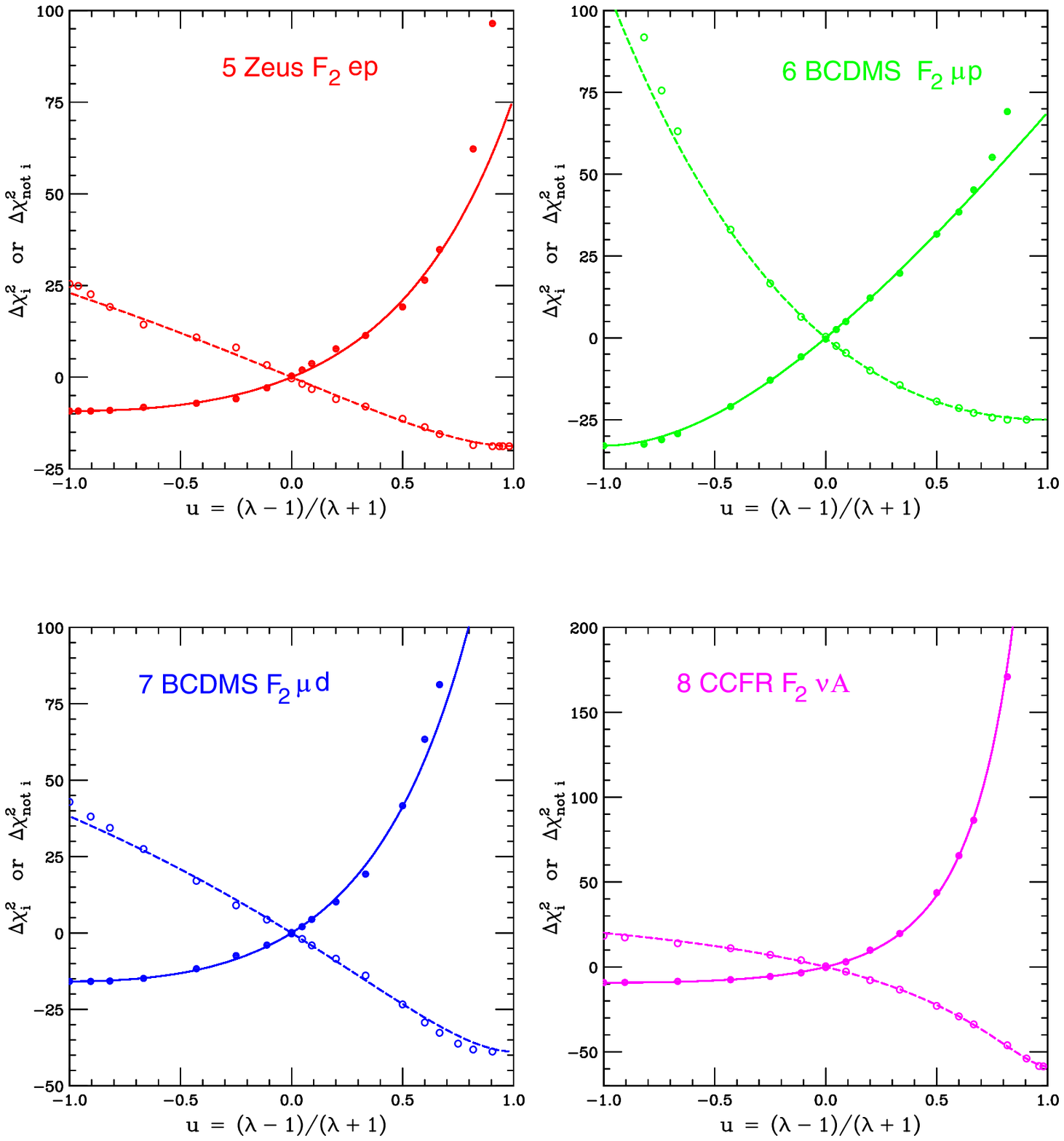}
  \caption{Variation of $\chi_i^2$ (solid curves) and $\chi_{{\rm not\,}i}^2$ 
   for experiments 5--8.
  }
  \label{fig:fit305678}
\end{figure}

%========================================================
\section{A third kind of plot}
\label{sec:ThirdKind}

The effective one-dimensional model of Appendix\ \ref{sec:one.param.model} 
suggests a new type of plot in which the various $\chi^2$ contributions 
are shown as a function of the Lagrange Multiplier parameter.
In doing this, it is convenient to define 
$u = (\lambda - 1)/(\lambda + 1)$.  This makes the function being minimized 
become $(1+u)\Delta_1 + (1-u)\Delta_2$, which is more symmetric with respect 
to $\Delta_1 = \chi_i^2$ and $\Delta_2 = \chi_{{\rm not\,}i}^2$.

Results from the 8 experiments discussed previously are shown in 
Figs.\ \ref{fig:fit301234} and \ref{fig:fit305678}.
The fits using the one-dimensional (2 parameter) model are quite good, 
which lends confidence to the results from that model that are shown 
in Table \ref{table:cteq}.

This fitting can be thought of as finding the single parameter (a non-linear 
combination of the fit parameters) that each experiment is most sensitive to 
in disagreement with the other experiments.

%========================================================
\section{Conclusions}

A common problem with global fits is that if there is a bad fit to a
particular experiment with few data points, its contribution to the total
$\chi^2$ may be completely negligible according to the hypothesis-testing
criterion.
The bad fit to a few data points will not be noticeable in the goodness of
fit as measured by the overall $\chi^2$.  Our new criterion applies a
parameter-fitting criterion to individual experiments, and hence can
recognize when a bad fit to a single experiment is significant, even
if that experiment has too few data points to make a big effect on the
total $\chi^2$.  A bad fit is detected if the experiment strongly
determines any particular combination of parameters, and that determination
is incompatible with the value determined by the other experiments.

In simple cases where there is only one parameter to measure, such as the mass 
of a particle, every one of a group of experiments can separately measure
the parameter.  Consistency between the experiments is just a matter of
whether the measured values agree within errors.  Of course, the new
criterion proposed in this paper reproduces that result; but the extra
complication it introduces is totally unnecessary for that situation.

The new criterion becomes important when there are many parameters to
determine, but each experiment determines only a few combinations of them.
Questions of consistency can then only be addressed after a global fit has
been performed to determine all of the parameters.  The more elaborate
methods that we propose then become essential to optimally test consistency.

Our criterion can be expressed in an especially simple form whenever the 
dependence of the individual $\chi_i^2$ can be approximated by a quadratic 
function of the effective parameters, as is shown in 
Appendix \ref{sec:one.param.model} for a single effective parameter, and in 
Appendix \ref{sec:quad.approx} for an arbitrary number of parameters.

We have demonstrated how our criterion works when applied to an actual
case of current interest, the global fit to determine parton densities.

%==========================================================
\section*{Acknowledgments}

This work was supported in part by the U.S.\ Department of Energy
under grant number DE--FG02--90ER--40577,
and by the National Science Foundation under grant number PHY--9802564.
JCC would like to thank the Alexander von Humboldt foundation for an
award.
JCC would like to thank W. Giele, L. Lyons, R. Thorne, and M.-J. Wang for
discussions.  
JP would like to thank W.-K. Tung and D. Stump for discussions.

%==========================================================
\appendix
\section{One-parameter quadratic model}
\label{sec:one.param.model}

In this appendix, we derive the relationship between $\chi_i^2$ and
$\chi_{\rm tot}^2$ for the simple case that experiment $i$ determines just
one combination of the fit parameters, and the dependence on that
parameter can be approximated by a quadratic function in the region of
$\chi_{\rm tot}^2$ that is of interest.  

We will show that in this case, a well-defined shape is
predicted for the graphs of $\chi_i^2$ vs.\ $\chi_{\rm tot}^2$, or
$\chi_i^2$ vs.\ $\chi_{\rm tot}^2 - \chi_i^2$.  This shape is defined
by two parameters; and the degree of consistency between experiment
$i$ and the other experiments is given directly by one of those parameters.
In Sec.\ \ref{sec:CTEQ5app}, we found this shape to provide a good
approximation for the study of fits of parton densities.

Since we are treating the case that experiment $i$ determines a single
combination of parameters, we can define a transformation of the
parameters, so that only a single parameter $p$ is relevant, and
\begin{eqnarray}
  \Delta \chi_{\rm tot}^2 &=& \chi_{\rm tot}^2 - \chi_{\rm tot}^2(0)
     \, = \, p^2  \\
  \Delta \chi_i^2 &=& \chi_i^2 - \chi_i^2(0)
     \, = \, (p^2 - 2pC)/D^2
\end{eqnarray}
where the argument $(0)$ denotes values at the minimum of $\chi_{\rm tot}^2$
as before.  The method of transforming $\Delta \chi_{\rm tot}^2$ and 
$\Delta \chi_i^2$ to this form follows from the argument given in Appendix
\ref{sec:quad.approx}.  There should be an extra term in 
$\Delta \chi_{\rm tot}^2$
quadratic in the other parameters, but for our calculation this extra term
is always set to its minimum value, and hence can be ignored.
Solving for the dependence of $\chi_i^2$ on $\chi_{\rm tot}^2$ leads to
\begin{equation}
\label{eq:tot.from.i}
   \Delta \chi_i^2 = \frac{ \Delta \chi_{\rm tot}^2 }{ D^2 }
           - \frac{ 2C \sqrt{ \Delta \chi_{\rm tot}^2 } }{ D^2 }.
\end{equation}

To interpret the two parameters $C$ and $D$ of this model, it is convenient
to express them as
\begin{eqnarray}
 \label{eq:CDdef}
  C &=& S / \! \tan \phi \nonumber \\
  D &=& 1 / \! \sin \phi \; .
\end{eqnarray}
Rescaling the fit parameter by $q = p \, \sin\phi \, \cos\phi$ then leads to
\begin{eqnarray}
  \Delta \chi_i^2 &=&
       \left(\frac{q}{\cos\phi} \, - \, S \cos\phi\right)^2 \,
       - \, (S \cos\phi)^2
       \label{eq:delchi1sq} \\
  \Delta\chi_{{\rm not\,}i}^2 = \Delta\chi_{\rm tot}^2 - \Delta \chi_i^2 &=&
       \left(\frac{q}{\sin\phi} \, + \, S \sin\phi \right)^2 \,
       - \, (S \sin\phi)^2 \; .
       \label{eq:delchi2sq}
\end{eqnarray}
It is easy to read off from these formulae that
experiment $i$ can be interpreted according to Gaussian statistics as a
measurement of $q$ with result
\begin{equation}
  q_1 = S \cos^2\phi \, \pm \, \cos\phi \; , \label{eq:psub1}
\end{equation}
while all of the other experiments combined give a result
\begin{equation}
  q_2 =  - S \sin^2\phi \, \pm \, \sin\phi \; . \label{eq:psub2}
\end{equation}
If we combine the errors in quadrature, these two results are seen to 
differ by
\begin{equation}
  q_1 - q_2 = S \, \pm \, 1 \; . \label{eq:psub1m2}
\end{equation}
Hence the two measurements differ by
\begin{equation}
  S = \sqrt{\frac{C^2}{D^2 - 1}} \label{eq:S}
\end{equation}
standard deviations.
They are therefore consistent with each other if and only if $S \lesssim 1$, 
if a Gaussian distribution of errors is assumed.
Meanwhile, the parameter $\tan \phi$ describes
how effective experiment $i$ is at measuring the parameter $q$, compared
to the combined effectiveness of the other experiments.

The relation between $\Delta \chi_i^2$ and $\Delta\chi_{{\rm not\,}i}^2$ in 
this model can be obtained explicitly by eliminating $q$ between 
Eqs.~(\ref{eq:delchi1sq}) and (\ref{eq:delchi2sq}).  The result is a
parabolic curve which is seen in Figs.\ \ref{fig:fit30a} and
\ref{fig:fit30b}: 
\begin{equation}
\label{eq:i.noti}
  \Delta_1 + \Delta_2 \, = \, 
  \left(\frac{\tan\phi \, \Delta_2 \, - \, \cot\phi \, \Delta_1}
             {2 \, S}\right)^2
\end{equation}
where $\Delta_1 \equiv \Delta\chi_i^2$ and 
$\Delta_2 \equiv \Delta\chi_{{\rm not\,}i}^2$.
This form holds in the region between the minimum of $\Delta\chi_i^2$ at 
$-(S \cos\phi)^2$ and the minimum of $\Delta\chi_{{\rm not\,}i}^2$ at
$-(S \sin\phi)^2$.
Outside that region, the curves do not exist.  

In a more general example, with more parameters, but where each experiment
only determines some of the parameters, the curves are no longer exactly
of the form of Eq.\ (\ref{eq:i.noti}).  This can happen both because there
are more parameters and because the quadratic approximation for the
$\chi^2$ functions may break down, particularly far from the global
minimum of $\chi_{\rm tot}^2$.  This presumably accounts for the
almost straight line behavior of the tails of some of the curves in Fig.\
\ref{fig:fit30b}.

%========================================================
\section{Quadratic approximation}
\label{sec:quad.approx}

It is instructive to assume that the various contributions $\chi_i^2$ which
make up $\chi_{\rm tot}^2$ can be approximated by quadratic functions of
the original fit parameters $\{x_i\}$ in the region of parameter space that
is allowed by the hypothesis-testing criterion for $\chi_{\rm tot}^2$.
This quadratic approximation has been found to be reasonably accurate in the
case of parton density fitting \cite{Pumplin:2001ct}; and in any case it can
provide a semi-quantitative guide to the kinds of behavior to be
expected.

The first derivatives of $\chi_{\rm tot}^2$ vanish at its minimum, so
in the quadratic approximation
\begin{equation}
  \Delta\chi_{\rm tot}^2 \, = \,
  \sum_{i,j} H_{ij}\, x_i \, x_j \; .
  \label{eq:appHxx}
\end{equation}
The real symmetric matrix $H$ has a complete orthonormal set of eigenvectors
$V_i^{(k)}$:
\begin{eqnarray}
  \sum_j H_{ij} \, V_j^{(k)} &=& \epsilon_k \, V_i^{(k)} \label{eq:appHV} \\
  \sum_j V_j^{(k)} \, V_j^{(\ell)} &=& \delta_{k \ell} \; . \label{eq:appVV}
\end{eqnarray}
Introducing new coordinates $\{y_i\}$ by
\begin{equation}
  x_i \, = \, \sum_k y_k \, V_i^{(k)} / \sqrt{\epsilon_k}
\end{equation}
leads to a simple diagonal expression:
\begin{equation}
  \Delta\chi_{\rm tot}^2 \, = \,
  \sum_{i} y_i^2 \; .
\end{equation}

The fit to a particular experiment, which we refer to here as experiment $1$,
is measured by $\chi_1^2$.  Since we assume that it is a quadratic function of
$\{x_i\}$, it is also a quadratic function of $\{y_i\}$:
\begin{equation}
  \Delta\chi_{1}^2 \, = \,
  \sum_{i} A_{i}\, y_i \, + \, \sum_{i,j} B_{ij} \, y_i \, y_j \; .
\end{equation}
Using arguments similar to the above, the real symmetric matrix $B$
has a complete set of orthonormal eigenvectors $W_i^{(k)}$:
\begin{eqnarray}
  \sum_j B_{ij} \, W_j^{(k)} &=& E_k \, W_i^{(k)} \label{eq:appBW} \\
  \sum_j W_j^{(k)} \, W_j^{(\ell)} &=& \delta_{k \ell} \; . \label{eq:appWW}
\end{eqnarray}
Introducing new coordinates $\{p_i\}$, this time without a change in scale, by
\begin{equation}
  y_i \, = \, \sum_k p_k \, W_i^{(k)}
\end{equation}
creates a diagonal form for $\Delta\chi_{1}^2$ while preserving the
very simple form for $\Delta\chi_{\rm tot}^2$:
\begin{eqnarray}
  \Delta\chi_{\rm tot}^2 \, &=& \, \sum_{i} p_i^2 \label{eq:appdelchitot} \\
  \Delta\chi_{1}^2 \, &=& \,
  \sum_{i} (\widetilde{A}_{i} \, p_i \, + \, E_{i} \, p_i^2) \; ,
  \label{eq:appdelchi1}
\end{eqnarray}
where $\widetilde{A}_{i} = \sum_j A_j W^{(i)}_j$.
This result is identical to the form that was {\it assumed} in 
Appendix\ \ref{sec:one.param.model}, except that there is a sum of independent
quadratic contributions in place of just a single one.

The relation between $\chi_1^2$ and $\chi_{\rm tot}^2$ under the quadratic
assumption is easily found by choosing the parameters $\{p_i\}$ so as to
minimize
\begin{equation}
  f_\omega = \omega \, \Delta\chi_1^2 \, + \, \Delta\chi_{\rm tot}^2 \; .
\end{equation}
The result is
\begin{equation}
  p_i = \frac{-\omega {\widetilde A}_i}{2 + 2 \omega E_i} \; .
 \label{eq:apppsubi}
\end{equation}
Substituting (\ref{eq:apppsubi}) into
(\ref{eq:appdelchitot})--(\ref{eq:appdelchi1}) relates  $\chi_1^2$ and
$\chi_{\rm tot}^2$ parametrically via the Lagrange multiplier parameter
$\omega = \lambda - 1$.

\clearpage

%========================================================


\begin{thebibliography}{99}

\bibitem{CTEQ}
%\cite{Lai:2000wy}
%\bibitem{Lai:2000wy}
H.L.~Lai {\it et al.}  [CTEQ Collaboration],
%``Global {QCD} analysis of parton structure of the nucleon: CTEQ5 parton
%distributions,''
Eur.\ Phys.\ J.\ {\bf C12}, 375 (2000)
[hep-ph/9903282].
%%CITATION = HEP-PH 9903282;%%

\bibitem{MRST}
%\cite{Martin:1998sq}
%\bibitem{Martin:1998sq}
A.D.~Martin, R.G.~Roberts, W.J.~Stirling and R.S.~Thorne,
%``Parton distributions: A new global analysis,''
Eur.\ Phys.\ J.\ {\bf C4}, 463 (1998)
[hep-ph/9803445].
%%CITATION = HEP-PH 9803445;%%

\bibitem{EW}
%\cite{Groom:2000in}
%\bibitem{Groom:2000in}
J. Erler and P. Langacker, ``Electroweak Model and Constraints on New
Physics'' in:
D.E.~Groom {\it et al.},
%``Review of particle physics,''
Eur.\ Phys.\ J.\ {\bf C15}, 1 (2000).
%%CITATION = EPHJA,C15,1;%%

%\cite{Giele:1998gw}
\bibitem{Giele:1998gw}
W.T.~Giele and S.~Keller,
%``Implications of hadron collider observables on parton distribution
%function uncertainties,''
Phys.\ Rev.\ D {\bf 58}, 094023 (1998)
[hep-ph/9803393].
%%CITATION = HEP-PH 9803393;%%

%\cite{Pumplin:2000vx}
\bibitem{Pumplin:2000vx}
J.~Pumplin, D.R.~Stump and W.-K.~Tung,
``Multivariate fitting and the error matrix in global analysis of data''
[hep-ph/0008191].
%%CITATION = HEP-PH 0008191;%%

%\cite{Stump:2001gu}
\bibitem{Stump:2001gu}
D.~Stump {\it et al.},
``Uncertainties of predictions from parton distribution functions I: the
Lagrange Multiplier method''
[hep-ph/0101051].
%%CITATION = HEP-PH 0101051;%%

%\cite{Pumplin:2001ct}
\bibitem{Pumplin:2001ct}
J.~Pumplin {\it et al.},
``Uncertainties of predictions from parton distribution functions II: the
Hessian method''
[hep-ph/0101032].
%%CITATION = HEP-PH 0101032;%%

\bibitem{paradox}
%\cite{Lyons:1984ar}
%\bibitem{Lyons:1984ar}
L.~Lyons,
``Parameter Fitting And Hypothesis Testing And Detailed Examples Of
Fitting Procedures,''
OXFORD-NP-94/84.

\bibitem{Bukhvostov}
%\cite{Bukhvostov:1997nf}
%\bibitem{Bukhvostov:1997nf}
A.P.~Bukhvostov,
``On the probability distribution of the experimental results''
[hep-ph/9705387].
%%CITATION = HEP-PH 9705387;%%

\bibitem{Pumplin}
  J. Pumplin, ``Systematic errors, non-Gaussian statistics, and effective
  chi-squared'', manuscript in preparation.

\end{thebibliography}
\end{document}